# Tuning the conductance of $H_2O@C_{60}$ by position of the encapsulated $H_2O$

Chengbo Zhu and Xiaolin Wang[a)]

*Spintronic and Electronic Materials Group, Institute for Superconducting and Electronic Materials, Australian Institute for Innovative Materials, University of Wollongong, North Wollongong, New South Wales 2500, Australia*

The change of conductance of single molecule junctions in response to various external stimuli is the fundamental mechanism for single-molecule electronic devices with multiple functionalities. We propose a concept that the conductance of molecule systems can be tuned from its inside. The conductance is varied in $C_{60}$ with encapsulated $H_2O$, $H_2O@C_{60}$. The transport properties of the $H_2O@C_{60}$-based nanostructure sandwiched between electrodes are studied using first-principles calculations based on the non-equilibrium Green's function formalism. Our results show that the conductance of the $H_2O@C_{60}$ is sensitive to the position of the $H_2O$ and its dipole direction inside the cage with changes in conductance up to 20%. Our study paves a way for the $H_2O@C_{60}$ molecule to be a new platform for novel molecule based electronics and sensors.

The emerging molecular electronics (MEs) based on single molecule offers a platform of miniaturization of devices, which are able to respond to various external excitations [1]. Thus, molecular electronic system is ideal for the study of charge transport at single molecule scale. [2-8] The drive to design functional molecular devices has pushed the study of metal-molecule-metal junction beyond the electronic transport characterization. [9] Single-molecule junctions have been investigated under a variety of physical stimuli, such as mechanical force, optical illumination and thermal gradients. In addition, spin- and quantum interference play import roles in ME. Various external stimuli that can affect the conductance of the ME systems are illustrated in Fig.1.

It should be pointed out that the change of conductance of single-molecule junctions in response to various external stimuli is the focus of the study in single-molecule electronic devices with multiple functionalities. It is well known that a system's electrical conductance or resistivity does not change unless variations are given to its shape, size and compositions through external influence. Here, we propose a concept that the conductance of molecule systems can be tuned from its inside, which renders a new degree of freedom for the changes of conductance without changing their physical appearance.

The systems, which have such effect, should be cavity-like and able to encapsulate objects with freedom of motion inside the cavity. This is absent in any classical materials. Besides, it breaks down for metallic cavity due to shielding effect. However, this type of systems and the effect could be possible in some single-molecule systems. It has come to our attention that the recently synthesized $H_2O$ encapsulated into $C_{60}$. $H_2O@C_{60}$ meets this criterion perfectly.

Encapsulating a single water molecule into the most common fullerene, $C_{60}$, has been accomplished experimentally [15]. The synthesized molecule, $H_2O@C_{60}$, is fascinating [16-19], as it provides a platform where the water molecule is isolated and prevented from forming any hydrogen bonding to other organic molecules or metals [17]. $H_2O@C_{60}$ is a remarkable molecule that consists of a polar molecule encapsulated into a highly symmetric and nonpolar cage. For $H_2O@C_{60}$, the polarity is no longer associated with its external shape. The encapsulated water molecule can rotate freely around the center inside the cage.

In this communication, the transport properties of the $H_2O@C_{60}$-based nanostructure sandwiched between electrodes are studied using first-principles calculations based on the non-equilibrium Green's function formalism. We demonstrate that, without changing the contact distance, the conductance of the $H_2O@C_{60}$-molecule junction is dependent on the position and the dipole direction of the encapsulated water molecule. Our study indicates that the $H_2O@C_{60}$ is a unique cage molecule for potential applications in molecular electronics and sensors.

The density functional theory (DFT)-based non-equilibrium Green's function (NEGF) formalism has been employed to calculate the transport properties [20]. We first relaxed the single molecule using SIESTA [21]. Then, the molecular junctions were constructed by placing the relaxed molecule between two Au tips in a 5×5 Au (111) unit cell, as shown in Fig. 2. The $H_2O@C_{60}$ molecule is connected to the electrodes with [6,6] double bonds [22]. The C-Au distance is set to 2.45 Å. The new structure is optimized again until the forces on all the $H_2O@C_{60}$ atoms are smaller than 0.03 eV/Å. The generalized gradient (GGA) Perdew-Burke-Ernzerhof (PBE) approximation is used for exchange-correlation [23]. Au atomic orbitals are described using single-zeta polarized orbitals, and molecular atoms are described by double-zeta polarized orbitals. The conductance transmission function is calculated using the Landauer formula [24]: $G(V_{bias}) = G_0 \sum T(V_{bias})$, where $T(E, V_{bias})$ is the transmission function at a given bias voltage $V_{bias}$, which is applied along the $z$ direction. $G_0 = 2e^2/h$ is the conductance quantum.

To see if the screening effect exists, we first examined the local currents between the encapsulated water molecule and the C atoms on the cage [25, 26], as shown in Fig. 3.

The red (blue) arrows represent the positive (negative) currents. It is obvious that there is a local current on the encapsulated water molecule, indicating that the Faraday cage disappears completely when the $H_2O@C_{60}$ molecule is sandwiched between electrodes under voltage bias. According to our calculations, the current flows mainly through the carbon bonds on the cage. There is still electron scattering from the C atoms to the water molecule, however, although it is very weak, being 1 per cent of the magnitude of the maximum current flowing between C bonds. As can be seen, all the positive currents first flow onto the O atom and then flow out of the water molecule from the two H atoms. The negative currents do the opposite: they first flow onto the two H atoms and then go through the O atom to the C atoms on the cage. Interestingly, the current paths are symmetrical with respect to the $y$-$z$ plane.

The Faraday cage is an enclosure formed by conducting materials. When the distance between the electrode and the fullerene molecule is shortened, the conductance increases rapidly [27]. We calculate the transmission when the $C_{60}$-Au distance is set to 3.2 a.u. The contact distance between the molecular edge and the surface of the electrode increases after relaxation. The junction is very conductive, and the conductance approaches 3.3 $G_0$. In such a highly conductive junction, the current still flows through the encapsulated water molecule. Therefore, the $C_{60}$ molecule cannot act as a Faraday cage when it is very conductive.

From our calculations, the gap for the $C_{60}$ molecule between the highest occupied molecular orbital (HOMO) and the lowest unoccupied molecular orbital (LUMO) is 1.65 eV, in agreement with Ref. [28]. The gap is slightly reduced to 1.62 eV by the encapsulation of the $H_2O$ molecule. The conductance of the $C_{60}$ junction and the $H_2O@C_{60}$ junction at zero bias is 0.592 $G_0$ and 0.577 $G_0$, respectively.

It is still controversial whether the encapsulated water molecule is able to move freely inside the cage [15, 18, 29]. Some believe that weak O-C coupling exists in the molecule [18]. In our calculations, when the $H_2O@C_{60}$ is bridged, the shortest O–C distance is 3.1686 Å, smaller than the summation of the van der Waals radii of the two atoms. The oxygen atom is 0.37 Å from the center of the bridged fullerene molecule after geometry optimization in our calculations. The dipole direction of the water molecule is almost along the $z$ direction.

We calculate the conductance and total energy for the $H_2O@C_{60}$ junction with the water molecule at different positions, as shown in Fig. 4. From the relaxed position, the water molecule is moved left 1.0 Å (L1.0), up 1.0 Å (U1.0), right 0.5 Å (R0.5), and right 1.0 Å (R1.0), while the dipole direction remains constant. Also, the conductance is calculated when the dipole direction is rotated 180 degrees around the x-axis after the encapsulated water molecule is moved 1.0 Å to the right (RR). We will refer to these possibilities as the L1.0-, U1.0-, R0.5-, R1.0-, and RR-junctions. During the calculation, the position of the $H_2O$ molecule is constrained. The conductances, their change ratios, and the total energies are plotted in Fig. 4(b). When the encapsulated water molecule moves right 0.5 Å, the distance between it and the center of the $C_{60}$ cage is shorter than that between its relaxed position and the center of the $C_{60}$ cage. It can be seen from Fig. 4(a) and (b) that when the water molecule moves toward the center of the $C_{60}$ cage, the conductance of the junction decreases.

Remarkably, our calculations demonstrate that the transport properties of the $H_2O@C_{60}$ molecular junction can be tuned by manipulating the encapsulated water molecule without changing the contact geometry. Also, the results show that the disappearance of the shielding effect is independent of the position of the water molecule.

As the water molecule moves further right to the position of R1.0, the conductance increases to 0.575 $G_0$, almost the same as for the $H_2O@C_{60}$ junction when the water molecule is at its relaxed position. Surprisingly, the conductance of the R1.0-junction increases when the dipole direction flips. As can be seen from Fig. 4(b), the total energy of the RR-junction is much lower than that of the R1.0-junction, suggesting that the water molecule would change its dipole direction if it moved to the position of R1.0. The water molecule does not necessarily change its dipole direction by 180 degrees, as only two dipole directions are calculated.

It is apparent that not only can the position of the molecule affect the conductance, but also the dipole direction of the water molecule can influence the conductance and the local currents. We therefore calculate the conductances and total energies for $H_2O@C_{60}$ junctions with dipoles of the encapsulated water molecule pointing in different directions, as shown in Fig. 4 (c) and (d). During the calculation, the oxygen atom is fixed at its relaxed position. Z, –Z, X, –X, Y, and –Y indicate the dipole direction of the water molecule. We will refer to these possibilities as Z-, –Z-, X-, –X-, Y-, and –Y-junctions. The Z-junction is the $H_2O@C_{60}$ junction with the water molecule at its relaxed position. As can be seen from Fig. 4(c), the conductance is clearly dependent on the dipole direction. When the dipole direction of the water molecule is along the –Z direction, the conductance is reduced. When the dipole points along Y or –Y, the conductance of the junction is larger. The total energy of the Y-junction is much higher than that of the –Y-junction. The conductances of the X-junction and –X-junction are both lower than that of the Z-junction. It is well known that the electrons of fullerene are reorganized with respect to the dipole direction in which the encapsulated $H_2O$ molecule points. The carbon atoms on the fullerene cage near the oxygen atom of the water molecule are slightly positively charged while those near the hydrogen atoms become slightly negatively charged [14, 15]. Thus, the conductance can be tuned by rotating the encapsulated water molecule.

There are many methods to tune the position and its orientation of $H_2O$ inside the cage such as light irradiation, magnetic and electric fields, heating, etc. All these external stimuli can 'communicate' with the water molecule causing it adjusting its location, which in turn changes the conductance of the $H_2O@C_{60}$ junctions. Our study paves a way for the $H_2O@C_{60}$ molecule to be a new platform for novel molecule based electronics and sensors.

In conclusion, we have theoretically investigated the transport properties of the single-molecule junction based on $H_2O@C_{60}$. The shielding effect disappears completely when the $H_2O@C_{60}$ molecule is sandwiched between electrodes. The disappearance of the shielding effect is independent of the position of the encapsulated water molecule. We demonstrate that, without changing the contact distance, the conductance of the $H_2O@C_{60}$-molecule junction is dependent on the position and the dipole direction of the encapsulated water molecule.

## ACKNOWLEDGEMENTS

X. L. Wang acknowledges support for this work from the Australian Research Council (ARC) through an ARC Discovery Project (DP130102956) and an ARC Professorial Future Fellowship Project (FT130100778). Computational resources used in this work were provided by Intersect Australia Ltd.

*To whom the correspondence should be addressed
   Email: xiaolin@uow.edu.au

1. M. Ratner, Nat. Nanotechnol. 8, 378 (2013).
2. M. A. Reed, C. Zhou, C. J. Muller, T. P. Burgin, and J. M. Tour, Science **278**, 252 (1997).
3. R. H. M. Smit, Y. Noat, C. Untiedt, N. D. Lang, M. C. van Hemert, and J. M. van Ruitenbeek, Nature **419**, 906 (2002).
4. B. Q. Xu, and N. J. Tao, Science **301**, 1221 (2003).
5. F. Chen, J. He, C. Nuckolls, T. Roberts, J. E. Klare, and S. Lindsay, Nano Lett. **5**, 503 (2005).
6. I. Díez-Pérez, J. Hihath, Y. Lee, L. Yu, L Adamska, M. A. Kozhushner, I. I. Oleynik, and N. Tao, Nature Chem. 1, 635–641 (2009).
7. H. Vazquez, R. Skouta, S. Schneebeli, M. Kamenetska, R. Breslow, L. Venkataraman, and M. Hybertsen, Nat. Nanotech.7 , 663 (2012).
8. S. Bilan, L. A. Zotti, F. Pauly, and J. C. Cuevas, Phys. Rev. B **85**, 205403 (2012).
9. S. V. Aradhya, L. Venkataraman, Nat. Nanotech. 2013, 8, 399.
10. N. Fournier, C. Wagner, C. Weiss, R. Temirov, and F. S. Tautz, Phys. Rev. B **84**, 035435 (2011).
11. Z. Liu, S. Y. Ding, Z. B. Chen, X. Wang, J. H. Tian, J. R. Anema, X. S. Zhou, D. Y. Wu, B. W. Mao, X. Xu, B. Ren, and Z. Q. Tian, Nat. Commun. **2**, 305 (2011).
12. E. A. Osorio, K. Moth-Poulsen, H. S. J. van der Zant, J. Paaske, P. Hedegaard, K. Flensberg, J. Bendix, and T. Børnholm, Nano Lett. **10**, 105 (2010).
13. S. K. Yee, J. A. Malen, A. Majumdar, and R. A. Segalman, Nano Lett. **11**, 4089 (2011).
14. C. M. Guédon, H. Valkenier, T. Markussen, K. S. Thygesen, J. C. Hummelen, and S. J. van der Molen, Nature Nanotech. **7**, 305 (2012)
15. K. Kurotobi and Y. Murata, Science **333**, 613 (2011).
16. B. Xu and X. Chen, Phys. Rev. Lett. **110**, 156103 (2013).
17. A. L. Balch, Science **333**, 531 (2011).
18. A. Varadwaj and P. R. Varadwaj, Chem.–Eur. J. **18**, 15345 (2012).
19. B. Ensing, F. Costanzo, and P. L. Silvestrelli, J. Phys. Chem. A **116**, 12184 (2012).
20. M. Brandbyge, J. L. Mozos, P. Ordejón, J. Taylor, and K. Stokbro, Phys. Rev. B **65**, 165401 (2002).
21. J. M. Soler, E. Artacho, J. D. Gale, A. García, J. Junquera, P. Ordejón, and D. Sánchez-Portal. J. Phys.: Cond. Mat. **14**, 2745 (2002).
22. T. Ono and K. Hirose, Phys. Rev. Lett. **98**, 026804 (2007).
23. J. P. Perdew, K. Burke, and M. Ernzerhof, Phys. Rev. Lett. **77**, 3865 (1996).
24. S. Datta, Electronic Transport in Mesoscopic Systems (Cambridge University Press, Cambridge, 1995).
25. M. Paulsson and M. Brandbyge, Phys. Rev. B **76**, 115117 (2007).
26. N. Okabayashi, M. Paulsson, H. Ueba, Y. Konda, and T. Komeda, Phys. Rev. Lett. **104**, 077801 (2010).
27. N. Néel, J. Kröger, L. Limot, T. Frederiksen, M. Brandbyge, and R. Berndt. Phys. Rev. Lett. **98**, 065502 (2007).
28. G. Géranton, C. Seiler, A. Bagrets, L. Venkataraman, and F. Evers, J. Chem. Phys. **139**, 234701 (2013).
29. A. B. Farimani, Y. Wu, and N. Aluru, Phys. Chem. Chem. Phys. **15**, 17993 (2013).

Figures and captions

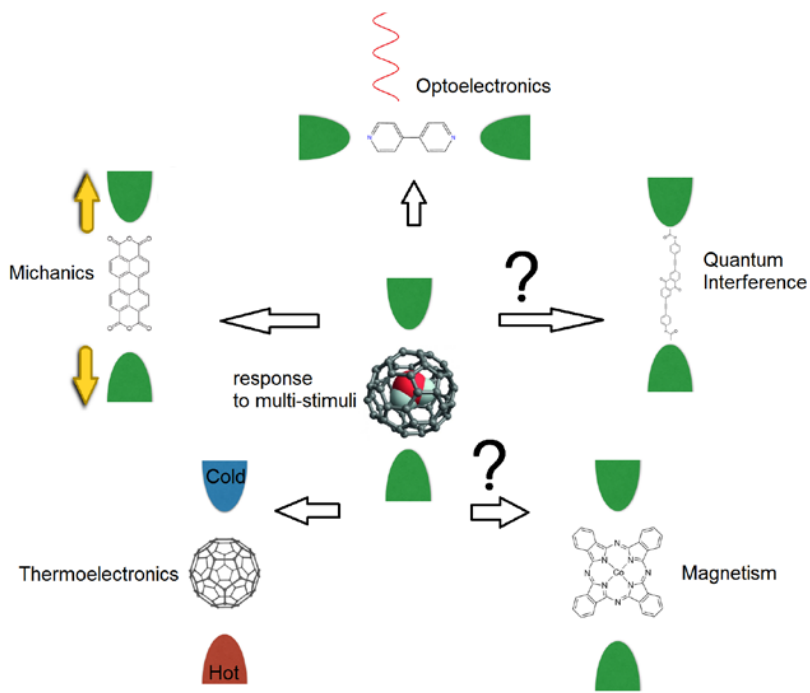

FIG.1 The change of conductance of single-molecule junction can be tuned from its inside, which may respond to various external stimuli [10-14].

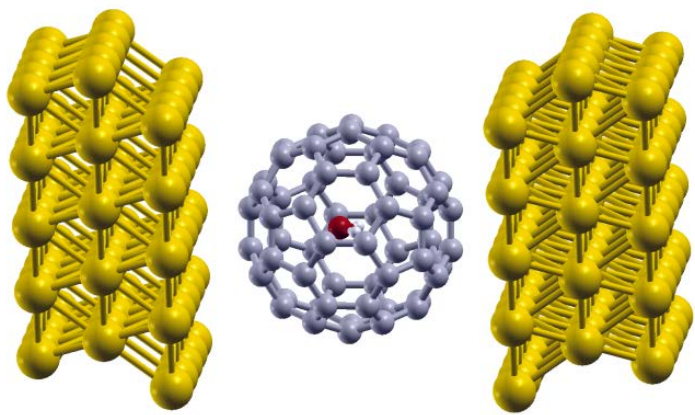

FIG. 2 Schematic illustration of the unit cell containing $H_2O@C_{60}$ used in the transport calculations. White atoms: H, grey: C, golden: Au.

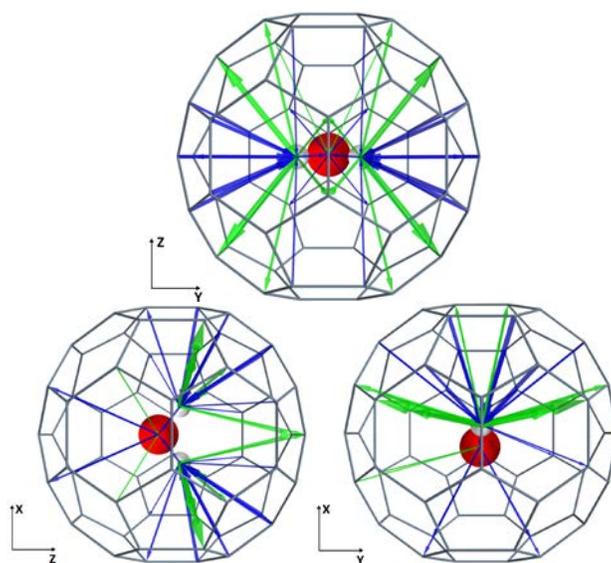
FIG. 3 Local currents between carbon atoms and water molecule, where the radius of the cylinder is proportional to the current density. Green currents represent the positive transport direction (along the $z$ direction), and blue currents represent the negative direction (along the $-z$ direction). The current is calculated at 0.5 V. It is obvious that the $C_{60}$ molecule cannot act as a Faraday cage because there are a number of current channels between the encapsulated water molecule and the C atoms.

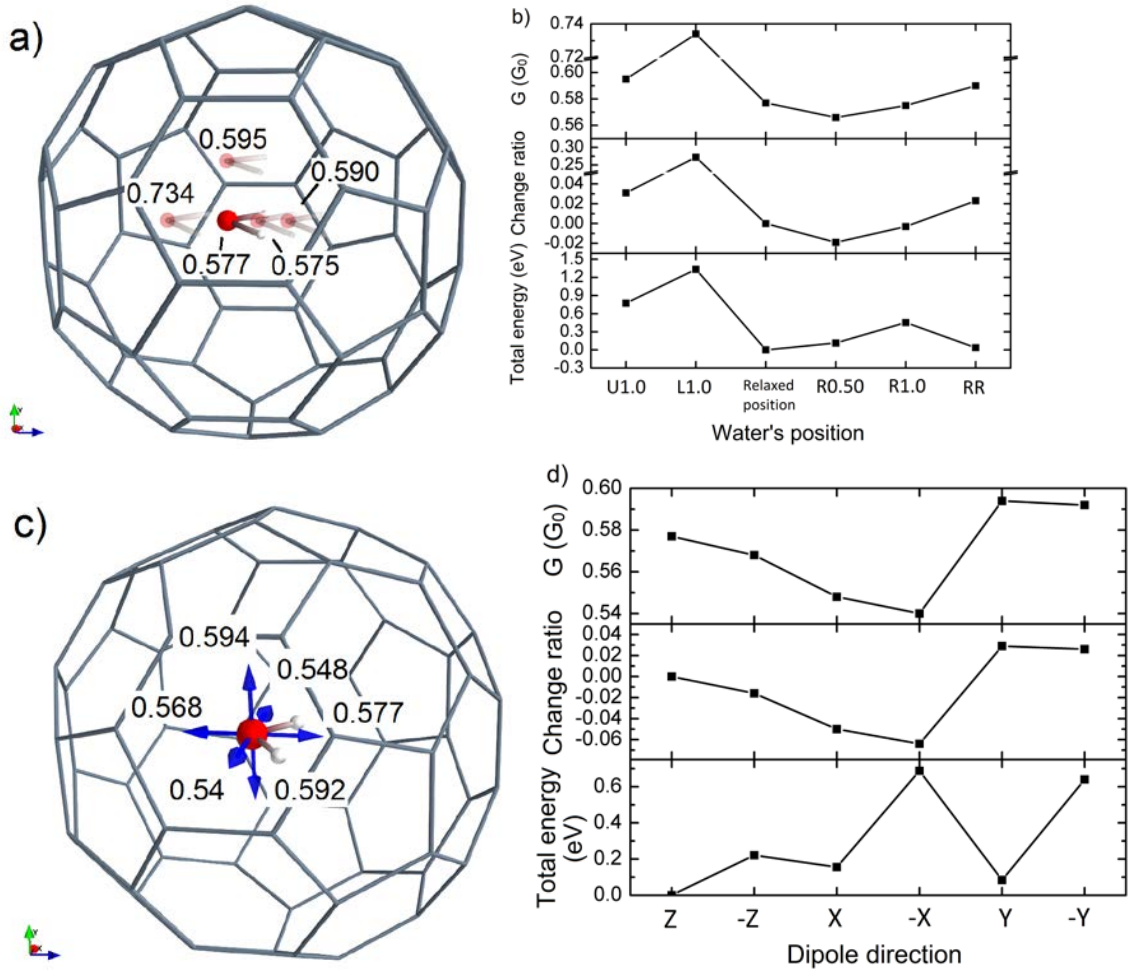

FIG. 4 a) and b) Conductance, its change ratio, and total energy for $H_2O@C_{60}$ junctions with the encapsulated water molecule at different positions; c) and d) Conductance, its change ratio, and total energy for $H_2O@C_{60}$ junctions with the dipole of the water molecule points in different directions. All conductance changes and total energies shown are relative to those of the $H_2O@C_{60}$ junction with the water molecule at the relaxed position. It is clear that, with the same contact geometry, the conductance is dependent not only on the position of the encapsulated water molecule, but also on the dipole direction of the water molecule inside.